\documentstyle[12pt]{article}
\newlength{\dinwidth}
\newlength{\dinmargin}
\newcommand{\ba}{\begin{array}}
\newcommand{\ea}{\end{array}}
\newcommand{\be}{\begin{equation}}
\newcommand{\ee}{\end{equation}}
\newcommand{\bea}{\begin{eqnarray}}
\newcommand{\eea}{\end{eqnarray}}
\newcommand{\gsim}{\mathrel{\mathop{\kern 0pt \rlap
  {\raise.2ex\hbox{$>$}}} \lower.9ex\hbox{\kern-.190em $\sim$}}}

\def\bbox{{\,\lower0.9pt\vbox{\hrule \hbox{\vrule height 0.2 cm
\hskip 0.2 cm \vrule height 0.2 cm}\hrule}\,}}
\newcommand{\dsl}{\pa \kern-0.5em /}

\newcommand{\nn}{\nonumber \\}
\def\* {&=&}

\def\eq#1{(\ref{#1})}

\begin{document}
\thispagestyle{empty}
\addtocounter{page}{-1}
\begin{flushright}
CERN-TH/2000-256\\
SNUST-000801\\
TIFR/TH/00-48\\
{\tt hep-th/0008214}
\end{flushright}
\vspace*{1.3cm}
\centerline{\large \bf A Note on D-Branes of Odd Codimensions}
\vskip0.3cm
\centerline{\large \bf {}From}
\vskip0.3cm
\centerline{\large \bf Noncommutative Tachyons
~\footnote{Work supported in part by BK-21 Initiative in Physics (SNU - 
Project 2), KRF International Collaboration Grant, KOSEF 
Interdisciplinary Research Grant 98-07-02-07-01-5, and KOSEF
Leading Scientist Grant.}}
\vspace*{1.5cm} 
\centerline{\bf Gautam Mandal ${}^{a,b}$ {\rm and} 
Soo-Jong Rey ${}^{a,c}$}
\vspace*{0.6cm}
\centerline{\it Theory Division, CERN, CH-1211, Genev\'e, Switzerland ${}^a$}
\vspace*{0.4cm}
\centerline{\it Tata Institute for Fundamental Research}
\vspace*{0.25cm}
\centerline{\it Homi Bhabha Road, Mumbai 400 005 India ${}^b$}
\vspace*{0.4cm}
\centerline{\it School of Physics \& Center for Theoretical Physics}
\vspace*{0.25cm}
\centerline{\it Seoul National University, Seoul 151-747 Korea ${}^c$}
\vspace*{1cm}
\centerline{\tt mandal@theory.tifr.res.in \hskip1.3cm sjrey@gravity.snu.ac.kr}
\vspace*{1.2cm}
\centerline{\bf abstract}
\vspace*{0.5cm}
On a noncommutative space of rank-1, we construct a codimension-one
soliton explicitly and, in the context of noncommutative bosonic open
string field theory, identify it with the D24-brane. We compute the
tension of the proposed D24-brane, yielding an exact value and show
that it is related to the tension of the codimension-two D23-brane by
the string T-duality.  This resolves an apparent puzzle posed by the
result of Harvey, Kraus, Larsen and Martinec and proves that the T-duality 
is a gauge symmetry; in particular, at strong
noncommutativity, it is part of the U($\infty$) gauge symmetry on
the worldvolume.  We also apply the result to non-BPS D-branes in
superstring theories and argue that the codimension-one soliton gives
rise to new descent relations among the non-BPS D-branes in Type IIA
and Type IIB string theories via T-duality.
\vspace*{.5cm}

\baselineskip=18pt
\newpage

%%%%%%%%%%%%%%%%%%%%%%%%%%%%%%%%%%%%%%%%%%%%%%%%%%%%%%%%%%%%%%%%%%%%%%%%%%%
%%%%%%%%%%%%%%%%%%%%%%%%%%%%%%%%%%%%%%%%%%%%%%%%%%%%%%%%%%%%%%%%%%%%%%%%%%%
In a series of recent important developments \cite{douglas, DH, SW},
it has been noted that noncommutative field theories in various
dimensions arise quite naturally at various corners of the moduli
space of nonperturbative string theories, especially, when a nonzero
value of the NS two-form potential $B_2$ is turned on, inducing
noncommutativity on the open string dynamics.

Among the most interesting developments are noncommutative solitons,
in particular, in the context of the string theories. At infinite
noncommutativity limit, Gopakumar, Minwalla and Strominger \cite{GMS}
have constructed explicit forms of static solitons by utilizing an
isomorphism between fields on noncommutative plane and operators on a
single-particle Hilbert space, ${\cal H}_\theta$ --- viz. the
Weyl-Moyal correspondence%
\footnote{Such solitons were constructed earlier in the
context of $c=1$ matrix model and two-dimensional QCD \cite{DMW}.}. 
The method of \cite{GMS} were adapted by Harvey, Kraus,
Larsen and Martinec \cite{HKLM} and Dasgupta, Mukhi and Rajesh
\cite{DMR} to bosonic string and superstring theories, respectively,
and have constructed a noncommutative version of D-branes, supporting
the viewpoint that D-branes are built out of solitonic lumps of
tachyon living on higher dimensional D-branes \cite{HK}. The result of
\cite{HKLM, DMR} has an important implication  for Sen's conjecture
\cite{sen} concerning universality of the tachyon potential in that,
in the limit of  infinitely strong  noncommutativity the level-zero
truncation yields {\sl exactly} Sen's conjectural relation between
the D-brane tension $T_{\rm D}$ and the height of the tachyon
potential ${ V}(T)$:
\bea
T_{\rm D} + \Big( { V}(0) - { V} (T_0) \Big) = 0.
\nonumber
\eea

In Ref. \cite{HKLM}, in the infinite noncommutativity limit, bosonic
D-branes were constructed explicitly only for those of even
codimensions, viz.  D23-, D21-, ... D1-branes but the rest, viz. D24-,
D22-, ..... D0-branes, were left out. In the commutative limit, all
D$p$-branes are on equal footing, as they are constructed out of the
tachyon field as localized energy lumps of both odd and even
codimensions. It is highly unlikely that, for some mysterious
reasons, the D-branes of odd codimensions are suppressed relative to
those of even codimensions as noncommutativity is turned on. How then
do the D-branes of odd codimensions arise in the infinite
noncommutativity limit? In this Letter, we point out that, in both
bosonic string and superstring theories, D-branes of odd codimensions
can be constructed by applying a U$(\infty)$ symmetry transformation.
It has already been emphasized in \cite{HKLM} that the
U$(\infty)$ symmetry arises as a gauge symmetry on the worldvolume of 
noncommutative D-branes. Moreover, the D-branes we will find are T-dual
to the ones found in \cite{HKLM}. As such, our result may be interpreted
as showing \footnote{We are grateful to J.A. Harvey for 
emphasizing to us the importance 
of the interpretation  and for further discussions.} 
that the T-duality, known to be a discrete (gauge) symmetry in compactified
bosonic and Type II superstring theories, is actually a part of the 
U$(\infty)$ gauge symmetry of the noncommutative D-branes.
 
We will begin with posing an issue concerning noncommutative
D$p$-branes in bosonic open string theory. In doing so, we will bring
up the puzzle concerning missing D$p$-branes for $p$=even more
explicit. At level-zero truncation, the bosonic open string field
theory is described by the action of the tachyon field $T$: 
\be S_0 =
{C \over G_{\rm st}} \int d^{26} y \sqrt{-G} \, \left( {1 \over 2}
G^{\mu \nu} F(T) \partial_\mu T \partial_\nu T + \cdots - { V}(T)
\right)_\star,
\label{sftaction}
\ee
in which the tachyon potential ${V}(T)$ is taken of the form
\bea
{V} (T) = - {1 \over 2} {\widetilde T}^2 + 
{1 \over 3 T_0}{\widetilde T}^3
\qquad {\rm where} \qquad
\widetilde{T}(x) = \exp \left[ \ln \left({3 \sqrt{3} \over 4} \right)
\partial_x^2 \right] T(x)
\label{potential}
\eea 
with a maximum at $T_0$ and a minimum at $T = 0$. The constant
$C$ is related to the tension of the D25-brane $T_{25}$ and the closed
string coupling parameter $g_{\rm st}$ as: \be C { {V}(T_0) \over
g_{\rm st} } = T_{25}, \nonumber \ee and $G_{\rm st}$, $G_{\mu \nu}$
refers to the open string coupling and metric, respectively.  The
$\star$-subscript refers to the fact that field products are defined
in terms of the noncommutative Moyal product: 
\be 
T(y) \star T(y) =
\exp \left({i \over 2} \theta^{\mu \nu} \partial_\mu^a \partial_\nu^b
\right) T(y_1) T(y_2) \Big\vert_{y_a, y_b \rightarrow y}.  
\nonumber
\ee 
Let us assume that noncommutativity is turned on along the ${\bf
y} = (y^1, y^2)$ directions only and has the strength $\theta^{12} =
\theta^{21} \equiv \theta$.  We shall be mostly interested in the
infinite noncommutativity limit: $\theta \rightarrow
\infty$. Rescaling the ${\bf y}$-coordinates along the
noncommutativity directions as ${\bf y} = \sqrt{\theta} {\bf x}$, both in
the quadratic gradient term and the cubic tachyon interaction term
Eq.(\ref{potential}), the action Eq.(\ref{sftaction}) 
can be recast as%
\footnote{Later, we will readdress the precise nature of the 
$\theta\to\infty$ limit.}
\be 
S_0 = {C \over G_{\rm
st}} \theta \int d^2{\bf x} d^{24} y \sqrt{-G} \, \left( {1 \over 2}
G^{\mu \nu} F(T) \partial_\mu T \partial_\nu T - {V}(T) + \cdots
\right)_\star,
\label{action-scaled}
\ee
where $\mu. \nu = 0, 3, 4, \cdots, 25$,  $V(T)$ refers to the tachyon 
potential in which $\widetilde{T}$ is simply set equal to $T$, and the 
ellipses denote subleading
terms of order ${\cal O}(1/\theta)$.
The $\star$-subscript now refers to the rescaled Moyal product
\bea
T({\bf x}) \star T({\bf x}) = \exp \left( {i \over 2} 
\left( \partial_1^a \partial_2^b - \partial_1^b \partial_2^a \right)\right)
T({\bf x}_a) T({\bf x}_b) \Big\vert_{{\bf x}_a, {\bf x}_b \rightarrow {\bf x}}.
\label{star}
\eea
The equation of motion is then given by 
\bea
G^{\mu \nu} \partial_\mu \partial_\nu T - {V}_\star'(T) = 0,
\nonumber 
\eea
which, in the static and translationally invariant case, reduces to
\bea
{V}_\star'(T) = - T + {1 \over T_0} T \star T = 0.
\nonumber
\label{seom}
\eea
In \cite{GMS}, a rotationally invariant static soliton 
satisfying Eq.(\ref{seom}) was found:
\bea
T({\bf x}) = T_0 \cdot 2 \exp \left( - {r^2 \over \theta} \right)
\qquad {\rm where} \qquad r^2 = y_1^2 + y_2^2. 
\label{rotsol}
\eea
 The tachyon field interpolates between $T = 2 T_0$,
{\sl twice} of the value of the potential maximum, and $T = 0$, the closed
string vacuum. 
The solution thus describes an object of $(23+1)$-dimensional
worldvolume, which can be interpreted \cite{HKLM} 
as a noncommutative version of the bosonic D23-brane. 

As shown in \cite{HKLM}, one can construct all other even 
codimensional D-branes by turning on
$B$-fields along $(x^1, x^2)$, $(x^3, x^4)$, ...., $(x^{24-2n}, x^{25-2n})$
subspaces. If the $B$-fields are all equal, then the resulting soliton
may be identified with a noncommutative version of the bosonic D$2n$-brane,
for which the ${\bf r}^2$ is replaced by ${\bf r}^2 = (x^1)^2 + \cdots
+ (x^{25-2n})^2$. Apparently, the noncommutative D-branes constructed this
way are only those with even codimensions. Where are the missing bosonic 
D-branes of odd codimensions? 

To answer the question posed, let us recapitulate the systematics of
constructing noncommutative solitons, following closely the works of
\cite{GMS} and \cite{HKLM}. The systematics is based on the so-called
Moyal-Weyl correspondence -- the equivalence  between the
configuration space of fields on a noncommutative
space whose algebra is defined
in terms of Moyal-product and the Hilbert space of Weyl-ordered
operators in one-particle quantum mechanics.

Consider a scalar field $T({\bf x})$ defined on  
a noncommutative plane with the associated Moyal-product 
Eq.(\ref{star}). The Fourier-mode function $\widetilde{T({\bf k})}$
is defined by the equation
\be
T({\bf x}) = \int {d^2 {\bf k} \over (2 \pi)^2}
\, \widetilde{T ({\bf k})} \, e^{ i {\bf k} \cdot {\bf x} },
\label{moyal}
\ee

Consider now operators $\widehat{x^1}, \widehat{x^2}$ 
satisfying the Heisenberg algebra
\bea
\left[ \widehat{x^1}, \widehat{x^2} \right] = i.
\nonumber
\eea
Built out of the operator algebra is an auxiliary Hilbert space 
${\cal H}_\theta$ of an auxiliary one-particle quantum mechanics. 
Given an operator $\widehat{T}(\widehat{\bf x})$ acting 
on ${\cal H}_\theta$,  one can define its Fourier-mode 
function as: 
\be
\widehat{T}(\widehat{\bf x}) 
= \int {d^2 {\bf k} \over (2 \pi)^2} \, \widetilde{\widehat{T} ({\bf k})} \, 
e^{ i {\bf k} \cdot \widehat{\bf x}}.
\label{weyl}
\ee
The Moyal-Weyl correspondence then refers to the equivalence%
\footnote{Refs. \cite{GMS,GN,AW} provide useful reviews of the formalism
in the context of noncommutative field theories and gauge theories.}  
between 
the $\star$-product algebra of  $T({\bf x})$ and the operator algebra of  
$\widehat{T}(\widehat{\bf x})$ under the condition that the Fourier mode
functions are set equal: 
\be
\widetilde{T({\bf k})} = 
\widetilde{\widehat{T}({\bf k})}
\label{equal-mode}
\ee It then follows straightforwardly
that
\bea
\widehat{T} (\widehat{\bf x}) \widehat{U} (\widehat{\bf x}) \quad
&\longleftrightarrow& \quad T ({\bf x} ) \star U ({\bf x})
\nonumber \\
{\rm Tr} \, \widehat{T} (\widehat{\bf x}) \widehat{U}
(\widehat{\bf x}) \cdots \quad & \longleftrightarrow & \quad
\int {d^2 {\bf x} \over (2 \pi \theta)} \, T({\bf x}) \star
U({\bf x}) \star \cdots.
\nonumber
\eea
In particular, the equivalence implies that static solitons are obtained
in the
operator formulation as distribution of  critical points of
the tachyon potential \footnote{
In the case of Eq.(\ref{potential}), the critical points consist of
$\{ 0, T_0 \}$.} among the set of one-dimensional projection operators in
${\cal H}_\theta$ --- an important observation made in 
Ref. \cite{GMS} and used in the context of
D-branes in \cite{HKLM}, \cite{DMR} and \cite{witten}. 

Stated as above, the Moyal-Weyl correspondence does not 
specify at all
what basis one needs to choose for the auxiliary Hilbert space 
${\cal H}_\theta$. 
In fact, no specific Hamiltonian of the analog one-particle
quantum mechanics is specified either. It is clear, however, that the
choice of the basis should reflect additional information on the phase
space labelled by ${\bf x}$ such as symmetries, presence/absence of 
boundaries, etc.  
As such, in what follows, we will consider two elementary 
choices of the basis -- simple harmonic oscillator basis and plane-wave 
basis -- and propose to identify them with D-branes of codimension even
and odd, respectively. 

As the first choice, let us take the energy eigenbasis 
$\{ \vert n \rangle, \,\, n = 0, 1, \cdots \}$ of a simple harmonic 
oscillator
as the `auxiliary' one-particle quantum mechanics. The basis is
built out of creation and annihilation operators:
\bea
a = {1 \over \sqrt{2}} \left( \lambda \widehat{x^1} + i {1 \over \lambda}
\widehat{x^2} \right) 
\qquad
a^\dagger = 
{1 \over \sqrt{2}} \left( \lambda \widehat{x^1} - i {1 \over \lambda}
\widehat{x^2} \right),
\label{def-lambda}
\eea
where $\lambda$ is a free parameter. They are defined by the following
action on the Hilbert space ${\cal H}_\theta$:
\bea 
a \vert n \rangle = \sqrt{n} \vert n + 1\rangle
\qquad a^\dagger \vert n \rangle = \sqrt{n + 1} \vert n + 1 \rangle
\nonumber
\eea
so that
\bea
\vert n \rangle = {a^{\dagger \, n} \over \sqrt{n!}} \vert 0 \rangle
\nonumber
\eea
and 
\bea
\vert m \rangle \langle n \vert = :
{a^{\dagger \, m} \over \sqrt{m!}} e^{ - a^\dagger a} {a^n \over 
\sqrt{n!}} : \, .
\nonumber
\eea
The most  general projection operator in the subspace 
spanned by the first $N$ eigenstates is given by
\bea
\widehat{\bf P} = \sum_{m,n=0}^{N-1} C_{mn} \vert m \rangle \langle n \vert
\qquad {\rm with} \qquad
\sum_n C_{mn} C_{nq} = C_{nq} \quad {\rm and} \quad
\sum_m C_{mm} = 1.
\nonumber
\eea
The conditions on $C_{mn}$'s are imposed 
to ensure the projection operator properties:
 $\widehat{\bf P}\cdot \widehat{\bf P} = \widehat{\bf P}$ and
Tr $\widehat{\bf P}$=1. 
As a simple choice with `zero' angular momentum
(that is, a diagonal projection operator), let us take
\bea
\widehat{\bf P}_{n} = \vert n \rangle \langle n \vert 
\quad \qquad (n = 0, 1, 2, \cdots).
\nonumber\eea

{}From Eq.(\ref{weyl}), one easily finds the corresponding Fourier 
mode-function $\widetilde{\widehat{P_n}}({\bf k})$ as:
\bea
\widetilde{\widehat{P_n}({\bf k})} = 
e^{ - { k}^2/4} L_n \left( {k}^2 / 2 \right),
\qquad 
\qquad \qquad k^2 = (k_1^2 + k_2^2),
\nonumber\eea
where $L_n$ denotes the $n$-th order Laguerre polynomial. 
Using \eq{equal-mode} and 
inserting the Fourier-mode function into Eq.(\ref{moyal}), one obtains
the Weyl-Moyal (inverse) map 
$\widetilde{\bf P}_n({\bf x})$ of the projection operator:
\be
{\bf P}_n({\bf x}) 
= 2 (-1)^n \, e^{ - \tilde{r}^2 } L_n (2 \tilde{r}^2 )
\qquad \,\,\, {\rm where} \qquad \,\,\,
\tilde{r}^2 = \left( \lambda^2 x_1^2 + {1 \over \lambda^2} x_2^2 \right).
\label{shosoliton}
\ee
{}From this the static noncommutative soliton of codimension-two can be 
constructed as $T({\bf x}) = T_0 \, {\bf P}_n ({\bf x})$.
One can check straightforwardly that ${\bf P}_n({\bf x})$ 
continues to satisfy the
properties of the projection operator: ${\bf P}_n({\bf x}) \star {\bf P}_n
({\bf x}) = {\bf P}_n ({\bf x})$ and 
$\int d^2 {\bf x} / (2 \pi \theta) \, {\bf P}_n ({\bf x}) = 1$.   
The solution Eq.(\ref{rotsol}) corresponds
to (a) the choice  of $n=0$ among all
possible codimension-two solitons in Eq.(\ref{shosoliton}) 
and (b) the rotationally 
symmetric choice of the $\lambda$-parameter, $\lambda = 1$. 

As the second choice, we will take projection
operators related to the `plane-wave' basis $\{ \vert p \rangle \}$ of the
Hilbert space  ${\cal H}_\theta$.  In ${\cal H}_\theta$, the `plane-wave' 
basis is defined by 
\bea
\widehat{x^2} \vert p \rangle  = p \vert p \rangle
\qquad \quad \langle q \vert p \rangle = {1 \over \sqrt{ 2 \pi}} e^{ i pq}
\nonumber \eea
and
\bea \langle p \vert p' \rangle = \delta (p, p')
\qquad \qquad  
\int {dp \over 2 \pi} \vert p \rangle \langle p \vert = {\bf I}.
\nonumber
\eea
The most general projection operator is given by
\bea
\widehat{\bf P} = \int {dp dp' \over (2 \pi)^2} C(p, p') \vert p \rangle
\langle p' \vert
\quad {\rm with} \quad
\int {dq \over 2 \pi} C(p, q) C(q, p') = C(p, p')
\quad {\rm and} \quad
\int {dp \over 2 \pi} C(p, p) = 1.
\nonumber
\eea
A simple choice for a `monochromatic' projection operator would be 
\bea
\widehat{\bf P}_{\widetilde{p_0}} = \vert \widetilde{p_0} \rangle \langle 
\widetilde{p_0} \vert.
\label{projection-plane-wave}
\eea
To be precise, $\vert \widetilde{p_0} \rangle$ is to be taken not as
a strict monochromatic wave of momentum $\widetilde{p_0}$ but as a wave-packet
built around mean-value of the momentum $\widetilde{p_0}$ and 
nonzero, finite variance. These conditions are necessary in order to
ensure $\langle \widetilde{p_0} | \widetilde{p_0} \rangle=1$
and to avoid complications associated with  delta-function (instead of
square-integrable) normalizability as well as subtle issues concerning infinite
noncommutativity and large deformation, which will be elaborated later. 
For definiteness, we may define $\vert \widetilde{p_0} \rangle$ 
operationally in terms of the following wave-packet:
\bea
\vert \widetilde{p_0} \rangle=
\int \frac{dp}{ \sigma\sqrt\pi} \exp \Big(-(p - \widetilde{p_0})^2/\sigma^2 
\Big) \vert p \rangle.
\nonumber
\eea
To obtain the monochromatic plane wave state, 
we let $\sigma \to 0$ in the end with a suitably defined limit procedure. 

It is straightforward to compute, as before, the Weyl-Moyal
(inverse) map 
${\bf P}_{p_0} ({\bf x})$ of the projection operator 
Eq.\eq{projection-plane-wave}. The result is
\be
{\bf P}_{\widetilde{p_0}} ({\bf x})= 2 \exp \left(- \frac{(x_2-
\widetilde{p_0})^2}{ \sigma^2} - \sigma^2 x_1^2 \right).
\label{soliton-plane-wave}
\ee
One easy finds that, as $\sigma \to 0$, the above function approaches 
\be
{\bf P}_{\widetilde{p_0}} ({\bf x}) = {1 \over 2\pi R_1} \delta(x_2 - 
\widetilde{p_0}),
\label{delta-shape}
\ee
consisting of a delta-function in $x^2$-direction while all dependence on
$x^1$-direction disappears. Here, $R_1$ is a suitably large radius of the 
$x^1$-direction and the factor $1/(2\pi R_1)$ is  to take care of the 
requisite normalization condition, where the `box cut-off' $R_1$
is related to the `Gaussian cut-off' 
$\sigma$ as $R_1 \sim 1/\sqrt \sigma$. 

{}From Eq.\eq{delta-shape}, one obtains a soliton of codimension-one as
$T({\bf x}) = T_0 \, {\bf P}_{\widetilde{p_0}} ({\bf x})$. The profile of this
soliton is very different from that of the circularly symmetric
soliton discussed previously.  Eq.(\ref{delta-shape}) represents, on
the noncommutative plane, a kink rather than a vortex. We will shortly
identify it as a D24-brane formed out of tachyon condensation on the
D25-brane worldvolume.
  
Before doing so, we should point out the underlying philosophy of our
construction. As mentioned before, various choices of orthonormal
bases of ${\cal H}_\theta$ are related to one another by U($\infty$)
transformations. As shown in \cite{HKLM}, the U($\infty$) transformations
are in fact gauge transformations on the D-brane worldvolume. 
A one-dimensional projection operator is given by $| \psi
\rangle \langle \psi |$ for any arbitrary state $| \psi \rangle$ in
${\cal H}_\theta$. What we found in the above construction is that the 
Weyl-Moyal map corresponding to different choices of $| \psi \rangle$ takes
generically  very different shapes on the noncommutative plane, and are
related to each other by  gauge transformations. 

In fact, different choices of the basis $| \psi \rangle$ are
characterized by the Grassmannian associated with a single soliton:
\bea
{\cal M}_1 \quad = \quad {{\rm U}(\infty) \over {\rm U} 
(\infty - 1) \times {\rm U}(1)}.
\nonumber
\eea
For example, the various values of $\lambda$ 
in Eq.\eq{def-lambda} correspond to a one-parameter
`squeezed state' subspace in the above manifold. 
The corresponding tachyon solutions are 
related to one another by U($\infty$) gauge transformations
\bea
T({\bf x}) \longrightarrow U({\bf x}) \star T({\bf x}) \star
U^\dagger ({\bf x}).
\nonumber
\eea
The U($\infty$) gauge transformations clearly respect the projection
operator properties: $({\bf P} \star {\bf P})({\bf x}) = {\bf P}({\bf
x})$ and $\int d {\bf x} / (2 \pi \theta) \, {\bf P}({\bf x}) = 1$;
they also preserve rank of the projection operator so that all points
of a gauge orbit correspond to the same rank.  Remarks similar to
these and the possibility of squeezing deformation have been made in
\cite{GMS,HKLM}; we have found here an explicit application of these
ideas to construct bosonic D24 branes.

As a concrete illustration of the U($\infty$) gauge transformation, we
now map the codimension-two profile Eq.(\ref{shosoliton}) to the
codimension-one profile Eq.(\ref{delta-shape}). This is achieved by
recalling that the parameter $\lambda$ is associated with the squeezing
deformation, an aspect discussed already in \cite{GMS}. One finds that
the $\lambda \rightarrow 0$ limit of Eq.(\ref{shosoliton}) reduces
exactly to Eq.(\ref{delta-shape}) for the choice $\widetilde{p_0} =
0$. In fact, in this limit, the squeezing deformation parameter
$\lambda$ plays precisely the same role as the regularization
parameter $\sigma$ in Eq.(\ref{soliton-plane-wave}). The opposite
squeezing limit $\lambda \rightarrow \infty$ yields again the same
state as Eq.(\ref{delta-shape}) but with $x^1 \rightarrow x^2, x^2
\rightarrow - x^1$ which amounts to a rotation in the noncommutative
plane. The moduli space of the squeezing deformation is parametrized
by $\lambda \in [0, 1]$. We have thus shown that the parameter $\lambda$
spans a one-parameter trajectory in ${\cal M}_1$ associated with the
squeezed state and the two extreme endpoint configurations represent
geometrically codimension-one as $\lambda \rightarrow 0$ and codimension-two 
as $\lambda \rightarrow 1$, respectively.

At this point, we would like to discuss an important subtlety %
\footnote{We would like to thank E. Witten for pointing out to us the
importance of this issue, which has already been discussed in
\cite{witten}.}  in applying the U($\infty$) gauge transformation and
obtaining the codimension-one, kink configuration. Recall that the
action Eq.(\ref{action-scaled})  was obtained by  dropping  terms
involving derivatives along the noncommutative directions ---
both in the 
quadratic gradient term and in the cubic interaction term --- by taking
$\theta \rightarrow \infty$ limit.  On the other hand, in applying the
squeezing deformation over the Grassmannian ${\cal M}_1$ and letting
$\lambda \rightarrow 0$, as described above, each derivative along the
noncommutative directions is amplified by a factor
$1/{\lambda}$. Certainly, to retain the utility of the infinite
noncommutativity and construction of the noncommutative D-branes via
the Moyal-Weyl correspondence, one ought to take a controlled
limit keeping $\theta_{\rm eff} \equiv \lambda^2 \theta$ fixed, while
letting $\theta \rightarrow \infty$ and $\lambda \rightarrow 0$.  In
order to suppress all the terms involving derivatives along the
noncommutative directions, it is necessary to take $\theta_{\rm eff}
\gg 1$.  In fact, one easily finds that, once the squeezing
deformation is taken, the new expansion parameter in
Eq.(\ref{action-scaled}) is set by $\theta_{\rm eff}$ instead of
$\theta$.  It is interesting to consider whether such subtleties 
involving large spatial derivatives arise in some other parts o
f the Grassmannian ${\cal M}_1$ as well.
We expect that the standard large-$\theta$ limit can be
taken  only in the {\sl interior} of the Grassmannian ${\cal M}_1$, 
but not along the boundary. As the boundary is approached, the 
large-$\theta$ constructions would in general require a suitably 
controlled scaling such as the one we have defined above.

Having obtained a codimension-one profile, 
we are naturally led to identify the soliton 
Eq.(\ref{delta-shape}) with a noncommutative version of the bosonic 
D24-brane. To ascertain this identification, we will now calculate the tension
of the codimension-one soliton and compare it with the known string theory 
result. For a static solution,
the action is simply the time interval times the mass, so the soliton mass 
can be equivalently calculated out of the action. We will calculate the 
action Eq.\eq{action-scaled} for the tachyon field
$T_{\rm sol} ({\bf x}, y) = T_0 \, {\bf P}_{\widetilde{p_0}} ({\bf x})$
where ${\bf P}_{p_0}({\bf x})$ is given by Eq.\eq{delta-shape}. 
To be specific, we will take the closed string background as
\bea
ds^2 &=& \theta(R_1^2 dx_1^2 +  dx_2^2) + dy_\mu dy^\mu
\nn
e^\phi &=& g_{\rm st}
\nn
B_2 &=& B \, dx^1 \wedge dx^2 \, .
\label{geometry}
\eea
{}From the definition  of the
open string coupling $G_{\rm st}$ and the open
string metric $G_{\mu\nu}$ \cite{SW}, one finds (cf. \cite{HKLM}) 
in   the large $\theta$ limit 
\be
G_{\rm st} = g_{\rm st}  \frac{\sqrt{G}}{2\pi \ell_{\rm st}^2 B}.
\ee
Using the results that  
\bea
{V}(T_{\rm sol}({\bf x}, y)) &=& {V}(T_0) \, {\bf P}_{\widetilde{p_0}} 
({\bf x})
\nn
\frac{C}{g_{\rm st}}{V}(T_0) &=& T_{25}
\nn  
\int {d^2 {\bf x}  \over 2 \pi \theta} \, {\bf P}_{\widetilde{p_0}} 
({\bf x}) &=& 1
\nn
1/B &=& \theta, 
\nonumber
\eea
we find that the action Eq.\eq{action-scaled} is evaluated to
\be
S= T_{25} (2 \pi \ell_{\rm st})^2 \int dt\ dy^3 \ldots dy^{25}. 
\label{action-value}
\ee
This agrees with the action of the codimension-two soliton evaluated 
in \cite{HKLM}, reflecting the U$(\infty)$ gauge invariance  
mentioned above, $\int d {\bf x} /(2 \pi \theta) \,\, U ({\bf x}) 
\star {\bf P}_{\widetilde{p_0}} ({\bf x}) \star U^\dagger({\bf x}) = 
\int d {\bf x} /(2 \pi \theta)\, \,{\bf P}_{\widetilde{p_0}} ({\bf x}) =
1$.

In \cite{HKLM}, the action Eq.\eq{action-value}
and the corresponding tension were interpreted  as those 
of a D23-brane extending along $y^3,\ldots, y^{25}$.
The symmetries of our codimension-one soliton are those of a 24-brane,
extending along $x^1, y^3, \ldots, y^{25}$ directions. We
will now show that, it is actually a D24-brane living in the
T-dual geometry where T-duality is performed
along the $x^1$ direction. The closed string geometry T-dual to 
Eq.(\ref{geometry}) is given  by
\bea
\widetilde{ds}^2 &=& \frac{1}{R_1^2} dx_1^2 - 
\frac{2 B}{R_1^2}dx_1 dx_2 
+ \left(1 + \frac{B^2}{R_1^2} \right) dx_2^2 + dy_\mu dy^\mu
\nn
e^{\widetilde{\phi}} &=& \widetilde{g_{\rm st}} = g_{\rm st} \ell_{\rm st}
/R_1
\nn
\widetilde{B} &=& B.
\label{dual-geometry}
\eea
Our assertion would follow if the worldvolume action of a D24-brane
extended along $x^1, y^3, \cdots, y^{25}$
in the T-dual background Eq.\eq{dual-geometry} is 
exactly the same as the action obtained from the tachyon soliton of 
codimension-one, Eq.\eq{action-value}. 
Begin with noting that the pull-back of the T-dual geometry on the
D24-brane worldvolume is given by
\bea
\widehat{ds}^2= \frac{1}{R_1^2} dx_1^2  + dy_\mu dy^\mu,
\nonumber
\eea
while pull-back of the $B$-field on the worldvolume 
vanishes. The worldvolume action of a D24-brane is therefore reduced to
\bea
S_{\rm DBI} = \widetilde{T_{24}} 
\int dt\ dx^1\ dx^{3}\ldots dx^{25} \sqrt{\widehat g},
\nonumber\eea
where $\widetilde{T_{24}}$ is the tension of a T-dualized D24-brane, given by
\bea
\widetilde{T_{24}}= T_{24} g_{\rm st} / \widetilde{g_{\rm st}} 
= T_{24} R_1/\ell_{\rm st}.
\nonumber\eea
Using the fact that $ \int dx^{1}\sqrt{\widehat g_{11}}= 2\pi /R_1$, we get
\be
S_{\rm DBI} = T_{24} (2\pi \ell_{\rm st}) \int dt\ dx^3 \ldots dx^{25}.
\label{action-d24}
\ee  
This is exactly  the same as Eq.\eq{action-value}, as
\bea
T_{24} = 2\pi \ell_{\rm st} T_{25}.
\nonumber
\eea
Hence, we have demonstrated that the 
codimension-one soliton Eq.(\ref{delta-shape}) is 
indeed the noncommutative version of D24-brane. Moreover, as in the D23-brane
case \cite{HKLM}, we have also found that the exact value of the D24-brane 
tension (T-dualized) has come out of level-zero truncation. 

Putting all the above together, we interpret the result as follows. The
noncommutative D24-brane on a circle we have constructed exhibits precisely 
the same properties as a noncommutative D23-brane distributed on a T-dual 
circle. As both D-branes have the same rank and are unitarily related 
to each 
other by (a subset of) U($\infty$) gauge transformation on ${\cal M}_1$, 
we have essentially shown that the T-duality, which is known to be a gauge
symmetry in bosonic string theory, is in fact a part of
U$(\infty$) gauge symmetry, at least in the large noncommutativity limit.   
While we have presented the noncommutative codimension-one solitons
for bosonic open string theory, the same analysis goes through for
noncommutative version of the non-BPS D-branes in Type II string
theories. It, however, entails an interesting variation of the
theme. The fact that {\em bosonic} D$p$-branes of both even and odd
$p$ constructed as above are related to each other by U($\infty$)
transformations does not contradict any known symmetry
principle. However, in superstring theory, for example Type IIA,
non-BPS $p$ branes arise only for odd $p$ (similarly for Type IIB only
even $p$). So, there clearly cannot exist a U$(\infty)$ symmetry
within either IIA or IIB superstring theory, which relates non-BPS
D-$p$ branes of different worldvolume dimensions. What, then, is the
interpretation of the U($\infty$) gauge transformation? We will now
show that the transformation relates non-BPS noncommutative D-branes
in Type IIA/IIB to those in Type IIB/IIA via the T-duality and hence
offers a new kind of D-brane descent relations.
 
Recall that in the commutative limit, Sen has given descent
relations among D-branes via worldvolume kinks or vortices formed out
of tachyon condensation.  The noncommutative D-brane descent relations
we will present are different from those of Sen's. Moreover, we will
find that the new series of descent relations provides a firmer
support to our interpretation regarding the U$(\infty)$ gauge transformation
as a T-duality between Type IIA and IIB strings through an analysis 
of the Chern-Simons coupling and Ramond-Ramond charges, an aspect which 
was not available for bosonic D-brane setup. 

Let us first recollect Sen's descent relations in the commutative
limit.  Begin with non-BPS D$2p$-branes in Type IIA string or
D$(2p-1)$-branes in Type IIB string. These non-BPS branes are
unstable, as represented by a tachyon field on the worldvolume.  The
tachyon is a real-valued field and it can condense in the form of a
topological kink on the worldvolume of the non-BPS D-brane.  As proven
by Sen, the kink turns out precisely the same as BPS D-brane.  Thus,
in Sen's descent relation, in Type IIA string, BPS D$(2p-1)$-brane can
be formed out of kink formation on the worldvolume of non-BPS
D$(2p)$-brane and, in Type IIB string, BPS D$(2p-2)$-brane can be
formed out of kink formation on the worldvolume of non-BPS
D$(2p-1)$-brane.

In infinitely strong noncommutative limit, solitons formed out of tachyon
condensation can come with two varieties: (1) codimension-two solitons 
studied in \cite{HKLM, DMR} or (2) codimension-one solitons proposed in the 
present paper. 

For the case (1), even though the solitons are objects of codimension-two, 
they ought to be identified with circular ring of codimension-one, BPS object.
This is because, as shown in \cite{DMR}, the Chern-Simons term on 
non-BPS D$p$-brane involves a coupling
\bea
{1 \over 2 T_0} \int_{M_{p+1}} dT \wedge C_p^{\rm RR},
\label{chern-simons}
\eea
where $C_p^{\rm RR}$ is the Ramond-Ramond $p$-form potential. Inserting
the noncommutative tachyon profile $T_{\rm sol} 
({\bf r}) = T_0 \, {\bf P}_0( r)$,
one finds that the above coupling yields
\bea
{1 \over 2 T_0} \int_{M_{p+1}} T_0 \, d {\bf P}_0 (r) \wedge C_p^{\rm RR}
\, \, = \,\, {1 \over 2} \int_{M_{p+1}} {\bf P}_0'(r) dr \wedge C_p^{\rm RR}.
\nonumber
\eea
Using the fact that $\int dr \, {\bf P}'_0(r) = - {\bf P}_0(0) = - 2$, one gets
precisely $-\int_{M_{p}} C_p^{\rm RR}$, implying that the soliton carries
(minus) one unit of the Ramond-Ramond charge, viz. BPS D$(p-1)$-brane. 
As the object is circularly
symmetric, a natural interpretation of the codimension-two soliton would
be that it is a ring of the BPS D$(p-1)$-brane on the noncommutative plane. 
Thus, one can say that the codimension-two soliton formation is nothing but 
noncommutative counterpart of the Sen's kink formation.  

For the case (2), we will now argue that the soliton ought to be interpreted
as the {\sl non}-BPS D$(p-1)$-brane of the T-dual, Type IIA/IIB string. 
To see this, let us evaluate the Chern-Simons coupling
\eq{chern-simons} for the codimension-one soliton 
profile $T_{\rm sol} ({\bf r}) =
 T_0 {\bf P}_{\widetilde{p_0}}({\bf x})$. We get
\bea
{1 \over 2 T_0} \int_{M_{p+1}} {T_0 \over R_1} 
d \delta (x^2) \wedge C_p^{\rm RR}
= {1 \over 2 R_1} \int_{M_{p+1}} \delta' (x^2) \wedge C_p^{\rm RR}.
\nonumber
\eea
In the present case, unlike the codimension-two solitons, 
$\int d x^2 \, \delta' (x^2) = 0$ and hence the above
Chern-Simons coupling
vanishes. This shows that the codimension-one soliton carries {\sl no} 
BPS$(p-1)$-brane charge at all. Clearly, this should
be considered as a non-BPS$(p-1)$-brane which does not
carry a RR charge. Furthermore such a non-BPS brane cannot
be in the original IIA/B theory since that contains
non-BPS $p$-branes. This implies that one must consider this
non-BPS $(p-1)$ as living in the T-dual geometry
(T-dualized along $x^1$), a fact that
we found in the bosonic context  by an explicit computation of 
the tension.  Furthermore, 
repeating the earlier analysis in the bosonic string context \cite{HKLM}, 
one discovers that tachyon fluctuation $\delta T (x^2)$ induces a coupling 
of  the form 
\bea
{1 \over 2 T_0} \int_{M_p} d \delta T \wedge C^{\rm RR}_{p-1}.
\nonumber
\eea
This is indeed the correct form of the Chern-Simons coupling for a non-BPS 
D$(p-1)$-brane localized at a point along the T-dualized 
$\widetilde{x^1}$-direction. Again, we have shown that the T-duality between 
Type IIA and IIB superstring theories, which is known as a discrete symmetry, 
is part of the U($\infty$) gauge symmetry arising in the large noncommutativity limit. 

We are grateful to M.R. Douglas and S.R. Wadia for useful discussions, and 
J.A. Harvey and E. Witten for important suggestions. We acknowledge warm 
hospitality of Theory group at CERN during this work.

\end{document}